\documentclass{aa}
\usepackage{graphics}
\newcommand{\be}{\begin{equation}}
\newcommand{\ee}{\end{equation}}
\begin{document}
%\thesaurus{11(05.01.1; 03.13.1; 03.13.2)}
\title{Extended radio emission after the 
soft X-ray maximum of the NOAA  
9077 AR solar flare on July 10, 2000}
\author{Shujuan Wang\inst{1,2}
\and Yihua Yan\inst{1,2} 
\and Qijun Fu\inst{1,2}}
\authorrunning{S.~J. Wang et al.}
\titlerunning{Extended radio emission after soft X-ray maximum}                 
\institute{Beijing Astronomical Observatory, Chinese Academy of Sciences, 
Beijing 100012, China
\and National Astronomical Observatories, 
Chinese Academy of Sciences, Beijing 100012, China}                     
\offprints{S.~J. Wang}
\mail{wsj@sun10.bao.ac.cn or wsj888@263.net}
\date{Received xxx/Accepted xxx}

%\begin{abstract}
\abstract{
An extended radio emission after a soft X-ray (SXR) maximum was detected 
in the active region NOAA 9077 by several observatories for the solar flare 
after 21:42 UT on July 10, 2000.
Also some radio fine structures before the enduring radio emission
were observed with the 1.0-2.0\,GHz spectrometer of Beijing Astronomical
Observatory (BAO) in the same time. We apply a shear-driven quadrupolar
reconnection model (SQR) to analyze the fine structures and the related radio
emission. We find that the footpoint shear motion of the flux loop
is accompanied with the emerging up of the loop during the reconnection 
process. We tentatively interpret the extended radio emission as 
the nonthermal radiation caused by a new reconnection process between
emerging flux loop and pre-existing overarching loop after the
soft X-ray maximum.
%\end{abstract}
}

\maketitle

\section{Introduction}
It is widely accepted that the nonthermal radio burst emission 
should take place {\it before\/} the thermal SXR maximum, for the basic 
process is thought to be the following: first, high-energy particles are 
accelerated during a reconnection of magnetic flux and propagate
along field lines, exciting the radio bursts; then, these  
particles pour down continuously, causing chromospheric
evaporation, and increasing the SXR emission to its peak (Heyvaerts
et al. 1977; Mclean \& Labrum 1985). 

However, an event of extended nonthermal emission {\it after\/}
the thermal maximum has been detected by several observatories
on July 10, 2000. This event seems
to suggest that the extended nonthermal radiation resulted from
repeated triggering of a new magnetic reconnection.
There are two different ways to interpret the triggering.
In the emerging flux model (Heyvaerts et al. 1977), when the magnetic
flux loops emerge from below the photosphere and interact with the
overlying field, continuous reconnection occurs
in the current sheet between the new and old fluxes. In the SQR model
(Aschwanden 1998), the shear motion of the footpoint will cause the
magnetic field between the sheared small-scale loop and the overlying
unsheared large-scale loop to become increasingly sheared and compressed,
triggering quadrupolar X-type reconnection. The latter interpretation also
regards, as a consequence of the reconnection, the various radio fine
structures, including type {I}{I}{I}, and U bursts, and reverse-slope
(RS)-drifting bursts. Indeed, some such radio fine structures were
observed with the 1.0-2.0\,GHz spectrometer of BAO in the event on
July 10, 2000.

\begin{table}
\caption[]{Data of the radio burst radiations and SXR burst}
\begin{tabular}{ccccccc}
\hline
\hline
Begin & Max  & End  & Obs. & Type &  Range     & Imp.    \\
%\scriptsize{(UT)} & \scriptsize{(UT)} & \scriptsize{(UT)}
% &      &      &\scriptsize{(MHz)}& (sfu) \\
(UT) & (UT) & (UT)
 &      &      &(MHz)& (sfu) \\
\hline
2121  & 2208 & 2314 & PALE  & RBR  & 245        & 13000   \\
2117  & 2205 & 2308 & PALE  & RBR  & 410        & 7200    \\
2117  & 2207 & 2254 & PALE  & RBR  & 606        & 1200    \\
2158E  & 2213 & 2249 & BAO  & RBR  & 2840       & 3085    \\
2117  & 2210 & 2257 & PALE  & RBR  & 15400      & 2600    \\
2105  & 2142 & 2227 &\scriptsize{GOES8}& XRB & 1-8{\AA} & M5.7   \\
\hline
\end{tabular}
\end{table}

This Letter first presents the observations of the event. 
Then we use the SQR model to analyze the radio fine structures and 
the related broadband radio bursts. We find that both the shear 
motion and emerging motion play a part during the magnetic reconnection
for this event.

\section{Observations}
The observations of PALE (Palehua station) and GOES 8 were prepared by
Space Environment Center (SEC) on the internet. The other observations
were made with the 1.0-2.0\,GHz solar radio fast dynamic spectrometer
and the 2.84\,GHz radio telescope at Huairou Solar Observing Station
of BAO (Fu et al. 1995). 

\begin{figure*}[t]
\resizebox{\hsize}{!}{\includegraphics{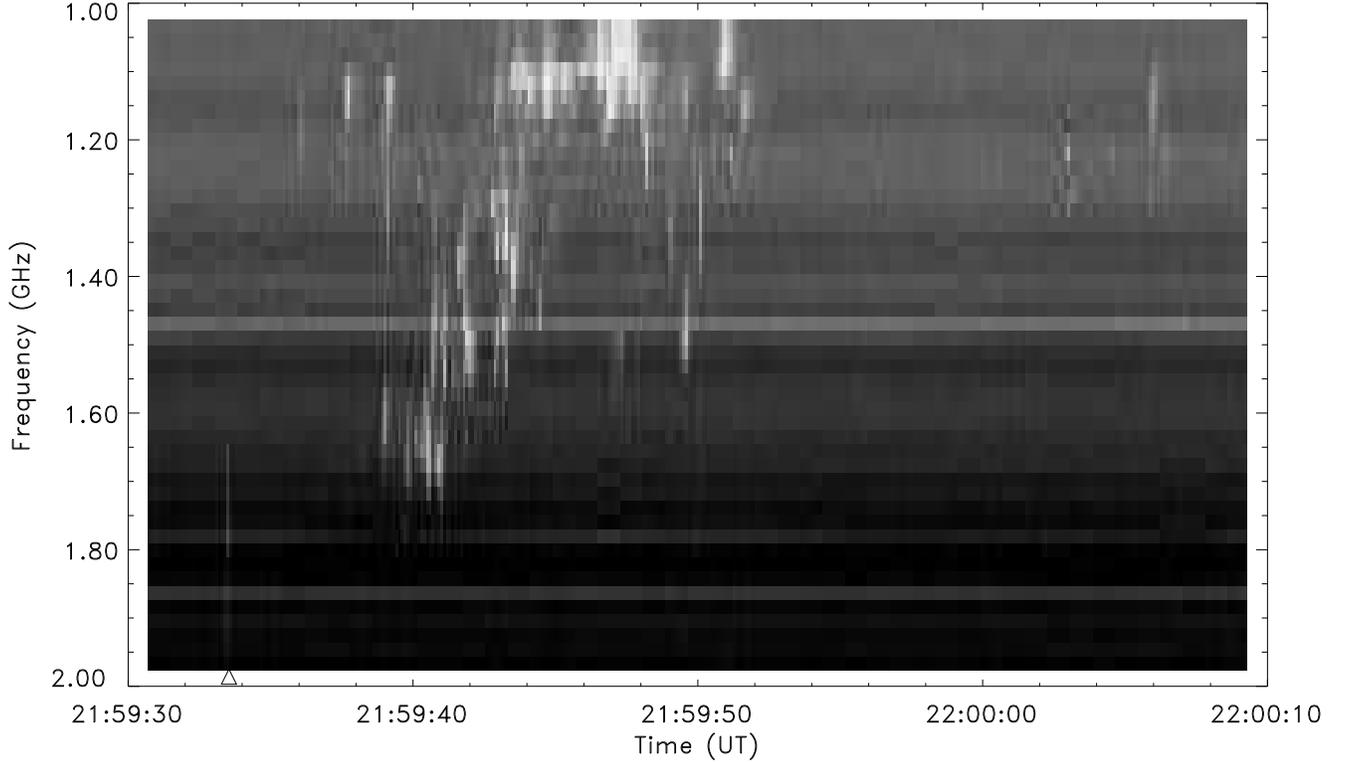}}
\caption{Gray-scale plot of the fine structures of the radio bursts 
on July 10, 2000 observed by BAO. Enhanced flux shown bright. 
The sign ``$\Delta$'' marks a very faint burst.}
\label{fig1} 
\end{figure*}

The July 10, 2000 flare, classified as M5.7/2B, was observed at 
21:05-22:27 UT in Active Region 9077 at position N19E49. Maximum
was at 21:42 UT. Morphologically, the flare was a two-ribbon flare.
The peak flux density at 2695 MHz was 140 sfu at 21:27 UT before 
the SXR maximum (by SEC). However an extended and very strong
radio emission (up to 3085 sfu at 2840 MHz) was detected after 21:42 UT,
and continued to about 23:00 UT, by several observatories. The data
are summarized in Table 1. It indicates repeated occurrence of magnetic
reconnection.

A group of radio fine structures before the broadband radio 
bursts were observed in the range of 1.0-2.0\,GHz 
21:59:30-22:00:10 UT (Fig.\,1). Fig.\,2 shows the GOES 8 
SXR data between 21:00 and 22:30 UT. Our radio data at the several
frequencies, which did not begin until 21:58 UT, are displayed in
the inset, and are further blown up in the grey-scale picture below.
The observed features can be summarized as follows.

(1) The fine structures contained four quasi-periods, the first one 
about 6\,s  between 21:59:32--21:59:38 UT, the second about 9\,s
between 21:59:38--21:59:47 UT, the third about 4\,s between 
21:59:47--21:59:52 UT, and the fourth about 14\,s between   
21:59:52--22:00:05 UT. The average period was about 6.5\,s. 

(2) Each period included both 
type U and type RS bursts. The type U bursts 
occurred in the lower frequency range, and the type RS bursts,
in the higher frequency range. All the type U bursts had a similar
turnover frequency of 1.16\,GHz.

(3) In each period, the flux density of the type U bursts was 3-4 
times stronger than the type RS bursts.

(4) The average frequency drift rate of the type U bursts was 
about 100 MHz s$^{-1}$ (both positive and negative), and that 
of the type RS bursts was about 
800 MHz s$^{-1}$ (negative). The former was clearly lower.

(5) The central frequencies of the type RS bursts drifted fast 
towards the lower frequency. The rate of drift decreased in time
in the range 100-40 MHz s$^{-1}$. 
 
(6) The type U bursts occurred after the type RS bursts.  

(7) At 22:04 UT, some 4 minutes after the fine structures, 
a broadband radio burst appeared, which lasted until 22:49 UT.  

\begin{figure}[t]
\resizebox{\hsize}{!}{\includegraphics{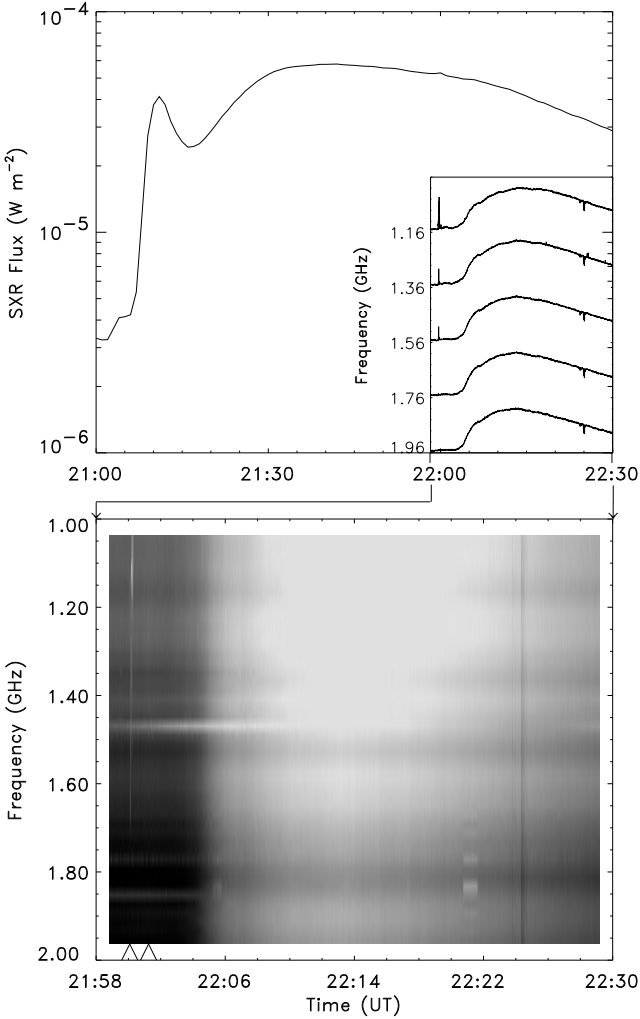}}
\caption{Comparison of the SXR emission with the related radio bursts.
{\em Top}: time profiles of the SXR flux (GOES 8) ({\em large}) between 
21:00-22:30 UT, and of the related radio emission ({\em small})
after 21:58 UT (no data being available before 21:58 UT in BAO).
{\em Bottom}: blow-up gray-scale plot of the radio emission 
in the range of 1.0-2.0\,GHz. 
The radio fine structures corresponding
to Fig.\,1 are the bright lines 
before 22:01 UT marked with double ``$\Delta$''.}
  \label{fig2} 
\end{figure}

\section{Theoretical analysis}
\subsection{SQR model}

According to Aschwanden (1998, Fig.\,1 of his paper),
the SQR model
may contain the following process.
Recognizing that 
shear motion along a neutral line to be an  
important condition for flaring,  
a consequence of a shear motion of the footpoint
along the neutral line is that the magnetic field
between the top of the
sheared small-scale loop and the overlying unsheared large-scale
field becomes increasingly sheared and compressed. At some point
the rising sheared loop will intersect with the unsheared large-scale
loop and trigger a quadrupolar X-type reconnection. During the 
reconnection, the connectivity of the field lines switches by 
exchanging the polarities of equal signs. The newly configured 
field lines then slip back from the X-point and relax to two 
dipole-like field lines. Then with the curvature of the
magnetic field reduced, magnetic energy will be released
(Aschwanden 1998, Aschwanden et al. 1999).

\subsection{Radio bursts}

As shown in Fig.\,3, upward propagating electron beams, 
along open and closed field lines, can be detected 
in the form of Type III, J and U bursts; while downward
propagating electron 
beams are detectable in the form of RS bursts. Therefore type U bursts
occur in the lower frequency range while type RS bursts occur in 
the higher. The following expected observational signature 
can also be seen in Fig.\,3:
a quasi-periodic sequence of type U bursts with similar
turnover frequencies produced by the upward propagating electron beams 
(Aschwanden 1998). The observational properties (1) and (2) 
in Sect.\,2 are in agreement with 
the theoretical predictions here. 

According to Zhao et al. (1997), 
if we assume the magnetic field
of the active region to be a dipole field above the photosphere,
then the relation between the height of radio source ($H$) and
its frequency ($f$) can
be estimated by

\begin{equation}
H=d\bigg[{\left(\frac{5.6B_0}{f~~(MHz)}\right)}^{1/3}-1\bigg].\\
\end{equation}
~\\
Where the depth of the dipole field
$d$ is about $3.5\times 10^4$ km, and $B_0$ is the magnetic 
field strength of the photosphere. For this event $B_0$ 
was about 2500\,G (Huairou Station, BAO).
Using the turnover frequency of 1.16\,GHz, the height of the top of 
the unsheared large-scale loop can be estimated to be 
about 4.5 $\times$ 10$^4$ km. 

\begin{figure*}[t]
\resizebox{\hsize}{!}{\includegraphics{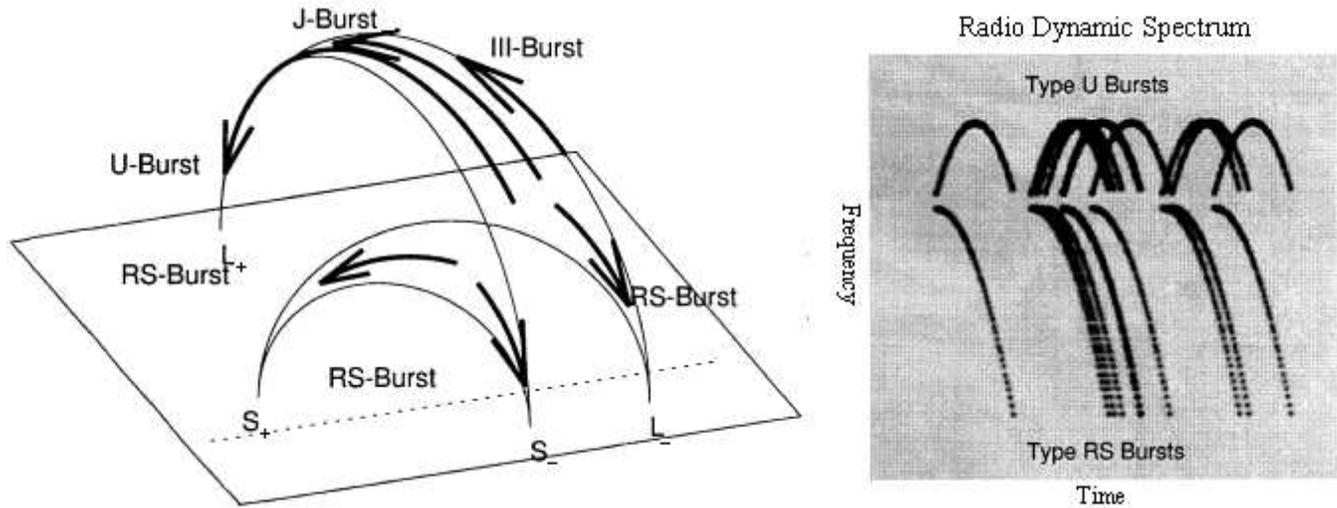}}
\caption{Spatial trajectories of radio bursts in the SQR model
(left frame) and the corresponding radio dynamic
spectrum (right frame). A quasi-periodic sequence of acceleration
episodes produces synchronized electron beams moving upwards (Type III,
J and U bursts) and downwards (RS bursts)
(Aschwanden 1998).}
  \label{fig3} 
\end{figure*}

Upward-accelerated electrons encounter a decreasing magnetic field,
which has a focusing effect on the pitch-angle distribution of the 
beam and makes it more prone to plasma emission. Also, the lower
density encountered in upward direction makes it easier for plasma
emission to escape. These two properties may be the main reason 
why upward propagating electron beams are detected much more frequently
(type U bursts etc.) than downward beams 
(RS bursts) (Aschwanden 1998). Therefore, when the upward and 
downward beams are both detected in radio, the flux density of type U 
bursts can be stronger than the related type RS bursts. This agrees
with  observational property (3) in Sect.\,2.

As the name implies, the large-scale loop have generally a much larger
size than the small-scale loop: the typical height is  
1.3 $\times$ 10$^5$ km for the former and 
2.0 $\times$ 10$^4$ km for the latter
(Aschwanden et al. 1992). The upward velocity of the 
electrons along the large-scale loop, then, is less than the 
downward velocity along the small-scale loop.
Thus, the frequency drift rate of type U bursts is 
lower than that of type RS, as was stated  under 
(4) in Sect.\,2. According to Eq.\,1, the spatial 
range ($\Delta H$) can be obtained for each of the type U  and
RS bursts. Dividing $\Delta H$ by the corresponding time interval,
we can obtain the electron velocity for each burst.  
The average upward 
and downward electron velocities were estimated to
be of 1.8 $\times$ 10$^3$ km s$^{-1}$ and 
1.4 $\times$ 10$^4$ km s$^{-1}$, respectively.

From the observational data, the central frequency 
(cf. the observed feature (5) in Sect.\,2) of each RS
burst can be obtained. Substituting the central frequency 
into Eq.\,1, the corresponding height ($H_c$) can 
be estimated for each RS burst.
We found $H_c$ increased in time.
This indicates 
that the small-scale loop was emerging from below. 
In the first period the velocity of emerging was
faster, but it gradually slowed down   
in the three succeeding periods (cf. Feature (1) in Sect.\,2).
Typical emerging velocity during the second period
is estimated at about 
1500 km s$^{-1}$. 
Thus, besides the shear
motion of the footpoint, which has been emphasized in the SQR model,
there is the emerging motion of small-scale magnetic loop taking place 
in the course of the reconnection.

It was also expected from the SQR model that for a large flare 
the reconnection will continue for a long time; 
for example, lasting more than 3 minutes for the 1994-Jan-06 
0405 flare (Aschwanden 1998). In this event, the flare was
a large flare (two-ribbon flare). 
Moreover the broadband radio bursts are believed 
to be the result of thermal bremsstrahlung from the hot and
dense plasma that remains in the low corona even for some tens
of minutes (Dulk 1985).
Thus we can understand the reason why 
the broadband radio bursts appeared after the fine structures
for about 4 minutes. 

\section{Discussion and Conclusions}
Several conclusions can be drawn from the present study:

(1) The extended radio emission was indeed caused by a new 
magnetic reconnection process.

(2) There was a flux loop emerging from the below. Typical 
emerging velocity was about 1500 km s$^{-1}$. 

(3) Either the shear motion or the emerging motion or both could have
triggered the quasi-periodic magnetic reconnection. 

(4) The radio fine structures provide clues for the understanding of
the process of energy release in the initial phase of the flare.

The original SQR model only considered the shear motion as the trigger
of the magnetic reconnection, but not the emerging motion.
However the emerging motion was indeed reflected in the present 
observations. The model may thus be improved in the future.
Besides, the type U bursts occurred after the type RS bursts, 
and the delay between the two tended to get shorter. 
This property may
have something to do with the emerging motion, because 
the consequence of the small-scale loop emerging is that
the tops of the two loops eventually reach the same height.

\begin{acknowledgements}
This research was supported by the National Nature Science
Foundation of China under grant No. 19773016 and No. 19973008,
and by Ministry of Science and Technology of China 
under grant No. G2000078403. 
We wish to thank the anonymous referee for advice on the 
improvement of the Letter. We are also grateful to Dr T.\,Kiang
for his revision on the language of the Letter.
\end{acknowledgements}


\begin{thebibliography}{}
\bibitem[1992]{aschwanden}
  Aschwanden, M.~J., Bastian, T.~S., Benz, A.~O., Brosius, J.~W.,
  1992, ApJ 391, 380 
\bibitem[1998]{}
  Aschwanden, M.~J., 1998, Solar Physics with Radio Observations,
  Proceedings of Nobeyama Symposium 1998, NRO Report 479, 307 
\bibitem[1999]{}
  Aschwanden, M.~J., Kosugi, T., Hanaoka, Y., et al., 
  1999, ApJ 526, 1026
\bibitem[1985]{dulk}
  Dulk, G.~A., 1985, ARA\&A 23, 169
\bibitem[1995]{fu}
  Fu, Q.~J., Qin, Z.~H., Ji, H.~R., et al., 1995, 
  Solar Phys. 160, 97
\bibitem[1977]{heyvaerts}
  Heyvaerts, J., Priest., E.~R., Rust, D.~M., 1977,
  ApJ 216, 123
\bibitem[1985]{mclean}
  Mclean, D.~J., Labrum, N.~R., 1985, Solar Radiophysics,
  Cambridge Univ. Press, Cambridge, UK
\bibitem[1997]{zhao}
  Zhao., R.~Y., Jin, S.~Z., Fu, Q.~J., 1997, Solar Radio 
  Microwave Bursts, Scientific Publishing House, Beijing, China
\end{thebibliography}
\end{document}